\documentclass[11pt]{article}
\usepackage{epsfig}
\usepackage{amsmath}
\usepackage{amsfonts}
\usepackage{amssymb}

\newtheorem{theorem}{Theorem}
\newtheorem{lemma}{Lemma}
\newtheorem{proposition}{Proposition}
\newtheorem{corollary}{Corollary}
\newtheorem{conjecture}{Conjecture}

\newenvironment{proof}{\noindent \emph{Proof.}\ }{\hfill
  $\Box$\vspace{1em}}
\def\constr{\mathop{\rm  constr}}
\def\Constr{\mathop{\rm  CONSTR}}
\def\opt{\mathop{\rm  opt}}
\def\op{\mathop{\rm  op}}

\def\Nset{\hbox{I\hskip-0.20em I\hskip-0.35em N}}
\def\Qset{\hbox{\hbox{Q\hskip-0.525em\lower-0.097ex
\hbox{\vrule height1.47ex width 0.07em}}\hskip0.50em}}
\def\Rset{\hbox{I\hskip-0.23em R}}

\begin{document}

\title{Optimizing diversity%
   \thanks{This research has been supported by the ADONET network of
     the European Community, which is a Marie Curie Training
     Network.}
}

\author{ Yannick Frein, Benjamin L\'ev\^eque, Andr\'as Seb\H{o}
  \\
  \\
  Laboratoire G-SCOP, INPG, UJF, CNRS, \\
  46, avenue Felix Viallet, 38031 Grenoble Cedex, France}

\maketitle

\begin{abstract}
  We consider the problem of minimizing the size of a family of sets $\mathcal G$ such that
  every subset of $\{1,\ldots,n\}$ can be written as a
  disjoint union of at most $k$ members of $\mathcal G$, where $k$ and $n$ are given numbers.
  This problem originates in a real-world application aiming at
  the diversity of industrial production. At the same time, the
  minimum of $|\mathcal G|$ so that every subset of $\{1,\ldots,n\}$
  is the union of two sets in $\mathcal G$
   has been asked
  by Erd\H{o}s and studied recently by F\"uredi and
  Katona without requiring the disjointness of the sets. A simple construction providing a feasible
  solution is conjectured  to be optimal for this problem for all values of $n$ and $k$ and regardless of the disjointness
  requirement; we prove this conjecture in
  special cases  including all $(n,k)$ for which $n\le 3k$ holds, and some individual values of $n$ and
  $k$.
  \


\end{abstract}

\section{Introduction}

The $n$-element set $\{1,\ldots,n\}$ is denoted by $[n]$. For two
positive integers $n,k$, a family $\mathcal G$ of subsets of $[n]$ is
said to {\em $k$-generate} $X\subseteq [n]$ if $X$ is the disjoint
union of at most $k$ members of $\mathcal G$. It $k$-generates the
family $\mathcal H\subseteq \mathcal P([n])$ if it $k$-generates every
$X\in\mathcal H$. It is called an $(n,k)$-generator if it generates
the entire powerset $\mathcal P ([n])$, that is, if every non-empty
subset of $[n]$ can be obtained as a disjoint union of at most $k$
members of $\mathcal G$. This work aims at determining the
$(n,k)$-generators of minimum size.  The {\em size} of a set is the
number of its elements (synonyme of cardinality).

Sets of size $1$ are called {\em singletons}. All the singletons
$\{i\}$ $(i=1,\ldots, n)$ must be contained in any $(n,k)$-generator.
We call an $(n,k)$-generator $\mathcal G$ of minimum size {\em
  optimal}, and introduce the notation $\opt(n,k):=|\mathcal G|$.  A
generator can be represented by a {\em hypergraph} (family of sets)
where the vertices are the elements of $[n]$ and the hyperedges are
the members of $\mathcal G$.

As Zolt\'an F\"uredi reports, Paul Erd\H{o}s \cite{Erd} asked about
the case $k=2$ allowing the target-sets to be {\em not necessarily
  disjoint} unions of two members of $\mathcal G$. He conjectured that
optimal generators consist of all the non-empty subsets of $V_1$ and
$V_2$, where $V_1,V_2$ is a partition of $[n]$ into two almost equal
parts. Since every subset of $[n]$ is the {\em disjoint} union of two
sets in this generator, it is implicit in this conjecture that the
optimum value does not depend on whether the two sets in the
definition are required to be disjoint or not.

Erd\H{o}s also considered the problem of generating only sets of size
at most $s$, where $s$ is a positive integer. F\"uredi and Katona
investigated this latter problem in \cite{FurKat}. For $s\leq 2$ the
problem is void, and for $s=3$ the problem is equivalent to Tur\'an's
theorem \cite{Tur}. For $s\le 4$, $n\geq 8$ they establish that the
cardinality of an optimal generator is
$n+(^n_2)-\lfloor\frac{4}{3}n\rfloor$.  When $s\le 4$ it does clearly
not matter whether the two sets are required to be disjoint or not.
(The same may be true for $s>4$ see Section~\ref{sec:construction},
but we cannot prove this.)  For all $s > 4$ the problem is apparently
open.

The same questions have been asked independently for optimizing the
diversity of production in the motorcar industry. To answer market
requirements, many companies want to reduce the delay between the
command and the delivery of a finished product, in the context of
offering a large choice for the possible options of these products.
The industrial problem that has to be faced is the following:
determine the \emph{semi-finished products} -- each of which
corresponds to a set of options -- that must be stocked in order to be
able to assemble any possible finished product in at most a given
number of operations \cite{Dac}. This latter constraint guarantees an
assembly time that does not exceed a desired time of delivery. The aim
is to minimize the size of the stock under this constraint. This is
equivalent to finding an optimal $(n,k)$-generator, where $n$ is the
number of options, and $k$ the maximum number of semi-finished
products that can be assembled.  From the viewpoint of industrial
technology the {\em disjointness} constraint cannot be relaxed, and it
is better to be able to generate {\em all subsets}.  Refining these
constraints, the optimization problems that can be stated occur to be
too difficult (NP-hard, see Section~\ref{sec:opt}); on the other hand,
these rigid requirements bring us to the prefixed constraints of
extremal combinatorics versus the flexible inputs of algorithmic
problems. These questions lead directly to beautiful and seemingly
difficult mathematical problems.

The basic problem studied in this article has been mentioned by the first
author in the activity report of the project ``decision making under
uncertainty'' at the Centre for Advanced Study of Oslo, in 2000-2001.
Conjecture~\ref{conj:Weak} below is explicitly mentioned in \cite{Dac}
independently of Erd\H{o}s \cite{Erd}.  However, the only result about
this problem so far seems to be \cite{FurKat}.

In Section~\ref{sec:construction} we introduce the main construction
and provide the related conjectures, remarks and some other
preliminaries, including the relation of the problem to the Tur\'an
number.  In Section~\ref{sec:proof} and Section~\ref{sec:main} the
main results of the paper and their proofs are presented, where
Section~\ref{sec:proof} is an auxiliary section collecting general
facts about the critical situation when for some $n,k,v$, $\mathcal G$
is not an optimal $(n,k)$ generator, but $\mathcal G-v$ is an optimal
$(n-1,k)$-generator. Finally, in Section~\ref{sec:opt} we show that
natural refinements of the problem in the spirit of combinatorial
optimization are NP-hard, and prove on the other hand that the
construction provides a generator that does never exceeds a small
constant times the optimum. In the Appendix we show some more results
concerning the case $k=2$, which enabled us to finish some more
concrete particular cases of the conjecture.

\section{Construction}\label{sec:construction}\label{sec:cons}

A natural way of constructing a generator is to partition the set
$[n]$ into $k$ parts and to include all the non-empty subsets of each
part in the generator.  The cardinality of such a generator is minimum
when the sizes of the parts differ by at most one.

More formally, let $p:=p(n,k):=\lceil \frac{n}{k} \rceil$ and
$r:=r(n,k)$ such that $n=p\, k-r$ with $0\le r<k$. Let
$V_1,\ldots,V_k$ be a partition of $[n]$ into $r$ sets of size $p-1$
and $k-r$ sets of size $p$. The generator we are constructing for all
$n,k\in\Nset$ is:
 $$\Constr(n,k):=(\mathcal P(V_1)\cup\cdots
 \cup\mathcal P(V_k))\backslash \{\emptyset\},$$ where $V$ is an
 arbitrary set. The cardinality of such a generator is
 $\constr(n,k):=r\times(2^{p-1}-1)+(k-r)\times(2^p-1)$. Note that
$$\constr(n,k) =\constr(n-1,k) +2^{p-1},$$ and  this simple
recursive formula  seems to be useful to keep in mind. It is
sufficient to prove {\em the same recursive formula for
$\opt(n,k)$.}

For instance we have $\constr(13,5)=27$ for $n=13$ and $k=5$.

Clearly, $\opt(n,k)\leq \constr(n,k)$, and in fact the equality seems
to hold always:

\begin{conjecture}
\label{conj:Weak} For all $n, k\in\Nset$ the generator
$\Constr(n,k)$ is optimal.
\end{conjecture}

Quite surprisingly this conjecture arose in production management, and
for $k=2$ it is a posthumus conjecture of Erd\H{o}s:

Indeed, as Zolt\'an F\"uredi reports, Erd\H{o}s \cite{Erd},
\cite{FurKat} asked the same question for $k=2$ without requiring the
disjointness of the sets. Could the same assertion be true for
arbitrary $k$ ?  Let $\op(n,k)$ denote the optimum for this problem.
Clearly, $\op(n,k)\le \opt(n,k)\le \constr(n,k)$, so if
$\op(n,k)=\constr(n,k)$ is true for some $(n,k)$, there is equality
throughout for this $(n,k)$.  These equalities would mean that
disjointness is an irrelevant requirement (in the sense that it does
not change the optimum value). Could this be proved by some simple
argument without necessarily settling the conjectures (see
Conjecture~\ref{conj:disjoint}) ?  In many results of the paper
$\opt(n,k)$ can be replaced by $\op(n,k)$, see some remarks at the end
of Section~\ref{sec:proof}.

Moreover, we also conjecture the unicity of the construction:

\begin{conjecture}
\label{conj:Strong}
 For  all $n,k\in\Nset$ such that  $p(n,k)\neq 2$,
 $\Constr(n,k)$ is the unique optimal $(n,k)$-generator.
\end{conjecture}

Trying to prove the preceding two conjectures inductively leads to the
following conjecture that would imply both (see the next section):

\medskip For a hypergraph $\mathcal G\subseteq\mathcal P([n])$ and
$z\in[n]$ let $\mathcal G(z):=\{g\in\mathcal G : z\in\mathcal G\}.$

\begin{conjecture}
  \label{conj:Degree} For all $n,k\in\Nset$, for every
  $(n,k)$-generator $\mathcal G$, there exists $z\in [n]$ such that
$$|\mathcal G(z)|\ge 2^{p(n,k)-1}.$$
\end{conjecture}

We prove that Conjecture \ref{conj:Degree} is true for $p=1,2,3$ and
$(n,k)\in\{(7,2),(8,2)\}$ for which $p=4$.

Notice that the partition underlying the construction is the same as
that in Tur\'an's theorem \cite{Tur}.  The two are actually related.
The Tur\'an number $T(n,s,l)$, where $n,s,l$ are three positive
integers with $l\leq s\leq n$, is the minimum number of subsets of
size $l$ of a set of size $n$, such that each subset of size $s$
contains at least one of them. In a generator, since every subset of
size $(l-1)k+1$ must contain a member of size at least $l$, there are
at least $T(n,(l-1)k+1,l)$ members of size at least $l$.

Tur\'an solved this problem for $l=2$.  If $l=2$, that is $s=k+1$, his
problem can be stated as follows: minimize the number of edges of a
graph on $n$ vertices so that the maximum number of pairwise
non-adjacent vertices does not exceed $k$. Replacing every member $g$
of a generator by a pair which is a subset of $g$, we always have this
property. Tur\'an proved that the unique minimum for this number is
given by $k$ cliques of almost equal size that partition the
vertex-set. This partition coincides with the defining partition of
the construction, showing that, the number of members of size at least
two in a generator is at least the number of sets of size exactly two
in Tur\'an's construction.

For $l\geq 3$, Tur\'an conjectured that the partition into blocks
still gives the solution to its problem, but this appears to be false.
According to Sidorenko \cite{Sid}, for $n=9$, $s=5$, $l=3$ with $k=2$
and $s=(l-1)k+1$ Tur\'an's construction provides
$\binom{4}{3}+\binom{5}{3}=14$ subsets of size $3$ so that every
$5$-tuple contains at least one of them, whereas the affine plane of
order $3$ gives a solution with only $12$ subsets with the same
property.  This example has been adopted by F\"uredi and Katona to
find the minimum number of sets that $2$-generate all $4$-tuples of a
set.

Indeed, for $n=9$, the set of minimum size that $2$-generates all
$4$-tuples can be defined with the help of the affine plane with
$q=3$: take the lines of two parallel classes ($6$ triplets) and the
$2$-element subsets of the lines for the two remaining parallel
classes ($9$ pairs for each, in total $18$). The generator $\mathcal
G$ consisting of these $24$ sets and the singletons $2$-generate all
the sets of size at most $4$.  Generalizing this construction F\"uredi
and Katona \cite{FurKat} prove that it provides the best estimate for
$2$-generating all $4$-tuples for all $n$. Compare $24$ with the size
of the subset of $\Constr(9,2)$ capable to achieve the same task, the
$2$- and $3$-tuples of $\Constr(9,2)$, $\binom{4}{3} + \binom{4}{2} +
\binom{5}{3} + \binom{5}{2}= 30$. With $30$ sets -- add to $\mathcal
G$ the $6$ lines of the affine space that are not yet included in it
-- actually the set of $5$-tuples can also be generated.

We cannot continue in this direction, since finding the Tur\'an number
when $l\geq 3$ is known as a difficult open problem, moreover a closer
direct look using more than just the containments provides better
lower bounds for the diversity problem in general
(Section~\ref{sec:opt}).

\section{Induction}\label{sec:proof}

In this section we show some general facts that may help in inductive
proofs provided we still have an optimal generator after the deletion
of one or two elements. In order to analyse how $\opt(n,k)$ changes as
a function of $n$ we need tight lower and upper estimates. The only
upper estimate we have is $\constr(n,k)$ and we will use it all the
time; in the lower estimates two parameters of a hypergraph will play
a role, the degree and the minimum transversal and the like:

\medskip For a hypergraph $\mathcal G \subseteq \mathcal P( [n])$ and
a subset $Z\subseteq [n]$ we define:
$$
\begin{array}{cclcc}
  \mathcal G-Z& :=&\{ g\in
  \mathcal G\ :\ g\cap Z =\emptyset
  \}\\
  \mathcal G(Z) & :=& \{ g\in \mathcal G\ :\ Z\subseteq g \}&&\\
  \mathcal G\sqcap Z& :=& \{g \cap Z\ :\ g\in\mathcal G\}&&\\
  \mathcal G\sqcup Z& :=& \{g \cup Z\ :\ g\in\mathcal G\}&&\\
  \mathcal G / Z &:=& \{ g \setminus Z\ :\ g\in \mathcal G\}
\end{array}
$$

One element sets $Z=\{z\}$ are often replaced by $z$, when the usage
is evident. Let us see some examples of occurrences of $z\in [n]$ and
$U\subseteq [n]$:
$$
\begin{array}{cclcc}
  \mathcal G - z& =&\{ g\in \mathcal G\ :\ z\notin g
  \}&=&\mathcal G\backslash\mathcal G(z)\\
  \mathcal G/z &=& \{ g\backslash\{z\}\ :\ g\in \mathcal G \} &\\
  \mathcal G(z)/z &=& \{ g\backslash\{z\}\ :\ g\in \mathcal G(z)\} &\\
  \mathcal G(z) - U & =&\{ g\in \mathcal G\ :\ z\in g,\, g\cap
  U=\emptyset \}&\\
  \mathcal G(z) \sqcup  U & =&\{g\cup U\ : g\in \mathcal G\ ,\ z\in
  g\}&
\end{array}
$$

The quantity $|\mathcal G(z) |$ is usually called the {\em degree} of
$z$ in the hypergraph $\mathcal G$. Note that $\mathcal
G(z)/z=\mathcal H$ if and only if $\mathcal G(z)=\{z\}\sqcup \mathcal
H$.

\medskip We will actually need to refine our sets and our quantities.
For a hypergraph $\mathcal G \subseteq \mathcal P( [n])$ and
$p\in\Nset$, $i=1,\ldots p,$ we denote $\,\,\,\mathcal G^i:=\{ g\in
\mathcal G\ :\ |g|\ge i\};\,\, \constr^i(n,k):=|\Constr^i(n,k)|.$

\medskip In $\Constr(13,5)$ there are $13$ hyperedges of size $1$,
$11$ of size $2$ and $3$ of size $3$, so $\constr^1(13,5)=27$,
$\constr^2(13,5)=14$, $\constr^3(13,5)=3$;
$\constr^i(n,k)-\constr^{i+1}(n,k)$ $(i=1,\ldots, p)$ is the number of
members of size exactly $i$.

We should not dream for anything stronger than
Conjecture~\ref{conj:Degree}, which implies already all the other
conjectures. However, we may need more details for a proof (as it will
be the case for some of our results):

\begin{conjecture}
 \label{conj:details}
 For all $n,k\in\Nset$, for every (n,k)-generator $\mathcal G$ we
 have:
$$|\mathcal G^i|\ge
\hbox{$\constr^i (n,k)$} \hbox{ for all } i=1,\ldots, p.\leqno{(1)}$$
\end{conjecture}

Since $\constr^1(n,k)=\constr(n,k)$ this conjecture contains
Conjecture~\ref{conj:Weak}.  When the average degree is not far from
the maximum (if $n=pk$ or more generally, when $r$ is small comparing
to $k$) it also implies Conjecture~\ref{conj:Degree}:

\begin{proposition}\label{prop:conjimpliesconj}
  If $n=pk$ and (1) holds for a hypergraph $\mathcal G$, then the
  average degree in $\mathcal G$ is at least $2^{p-1}$, and every
  degree is equal to this number if and only if there is equality
  everywhere in (1).
\end{proposition}

\begin{proof} The average degree of $\mathcal G$ is equal to the sum
  of the sizes in $\mathcal G$ divided by $n$, which in turn is equal
  to
$$1/n\sum_{i=1}^n |\mathcal G^i|,$$
since a set of size $s$ is encountered here for the values $i=1,
\ldots, s$, that is, exactly $s$ times.

If (1) holds, then this number is greater than or equal to the average
degree of the hypergraph $\Constr(n,k)$, which is equal to $2^{p-1}$,
since all degrees are equal to this number. Therefore all degrees are
equal to $2^{p-1}$ if and only if there is equality everywhere in (1),
as claimed. \end{proof}

\begin{proposition}\label{prop:formula} For all $i=1,\ldots,p:|\{g\in\Constr(n,k): |g|=i\}|=$
$$\hbox{$\constr^i(n,k)$} - \hbox{$\constr^{i+1}$}(n,k)=r{p-1\choose
  i}+(k-r){p\choose i}.$$
\end{proposition} {\hfill $\Box$\vspace{1em}}

If $H\subseteq \mathcal P([n])$ is a hypergraph, a {\em transversal}
is a set that meets all members of $\mathcal H$, and $\tau (\mathcal
H)$ denotes the minimum size of a transversal. If $\mathcal H$ has $m$
disjoint members, then clearly $\tau (\mathcal H)\ge m$. If $\mathcal
H$ contains the empty set, it has no transversal, we define then $\tau
(\mathcal H)=\infty.$

Generators can be characterized in term of transversals, by the
following easy but useful proposition:

\begin{proposition}\label{prop:tau} Let $\mathcal G\subseteq \mathcal
  P([n])$ be an $(n,k)$-generator,
  and $i\in\{1,\ldots, p\}$. Then $\tau (\mathcal G^i)\ge k(p-i+1)-r,$
  and this bound is tight.
\end{proposition}

\begin{proof} Suppose $\mathcal G$ is an $(n,k)$-generator, and
  $T\subseteq [n]$, $|T|< k(p-i+1)-r$. Then
  $|V-T|=n-|T|>kp-r-(k(p-i+1)-r)=k(i-1)$, so in a partition into $k$
  elements there is a part of size at least $i$, so $T$ is not a
  transversal of $\mathcal G^i$, and the proposition is proved. The
  equality holds for $\mathcal G=\Constr(n,k)$.
\end{proof}

The extreme case $i=p$ of Conjecture~\ref{conj:details} is now easy,
and we will need it:
\begin{proposition}\label{prop:extreme} If $\mathcal G$ is an
$(n,k)$-generator, then $|\mathcal G^p|\ge
\constr^p(n,k)=k-r=n-(p-1)k,$ and if the equality holds $\mathcal G$
contains exactly $k-r$ sets of size at least $p$, and they are
pairwise disjoint.
\end{proposition}
\begin{proof} Apply the preceding proposition to $i=p$: $|\mathcal
G^p|\ge\tau(\mathcal G^p)\ge k-r= n-(p-1)k$, and if the equality
holds throughout, then in particular $|\mathcal G^p|=\tau(\mathcal
G^p)$, that is, all the sets of $\mathcal G^p$ are pairwise
disjoint.
\end{proof}

\bigskip We prove now that {\em Conjecture~\ref{conj:Degree} implies
  Conjecture~\ref{conj:Weak}, and Conjecture~\ref{conj:Strong}.}

The following lemma deduces the optimality of the construction -- that
is, Conjecture~\ref{conj:Weak} -- by induction on $n$ if and only if
there always exists an optimal generator containing a vertex of degree
at least $2^{p-1}$ (which is somewhat weaker than
Conjecture~\ref{conj:Degree}, see Conjecture~\ref{conj:WeakDegree}
below.):

\begin{lemma}
  \label{lemma:WeakInduct}
 Let $\mathcal G$ be an
  optimal $(n,k)$-generator, $z\in [n]$, $|\mathcal G(z)|\ge 2^{p-1}$,
  and  assume $\constr(n-1,k)=\opt(n-1,k)$. Then:
$$|\mathcal G(z)|= 2^{p-1},\,\,\constr(n,k)=\opt(n,k).$$
\end{lemma}

\begin{proof}
  Since $\mathcal G - z$ generates $\mathcal P([n]\setminus\{z\})$, it
  is an $(n-1,k)$-generator:
  $$\opt(n,k)=|\mathcal G|= |\mathcal G(z)| + |\mathcal G-z|\ge 2^{p-1}+
  \opt(n-1,k)$$$$=2^{p-1}+ \constr(n-1,k)=\constr(n,k)$$ so there is
  equality everywhere.
\end{proof}

As a consequence, we see that Conjecture~\ref{conj:Weak} follows
recursively for $(n,k)$ if we know Conjecture~\ref{conj:Degree} for
all $(n',k)$, $k\le n'< n$.

\medskip

This recursion raises the question of analysing ``the moment when a
generator deviates from the construction, while $n$ is increased and
$k$ is fixed''. (We will see that Conjecture~\ref{conj:Degree} is true
if $n\le 3k)$. In the construction there are vertices $z$ for which
$\Constr(n,k)-z$ is isomorphic to $\Constr(n-1,k)$. The following
theorem shows that $|\mathcal G(z)|$ with $\mathcal
G-z=\Constr(n-1,k)$ has to pay a ``high price'' for essentially
deviating from the construction:

\bigskip If $\mathcal H$ is a hypergraph on $[n]$, $z\in [n]$ and
$z\notin U\subseteq [n]$, we say that $z$ {\em sees} $U$ if $\mathcal
G(z)\sqcap U =\mathcal P (U)$. Furthermore it {\em strongly sees} $U$
if $\mathcal G(z)\supseteq \{z\}\sqcup \mathcal P (U)$.

\begin{theorem}
  \label{thm:Strong}
  Let $\mathcal G\subseteq \mathcal P ([n])$
   be an $(n,k)$-generator, $z\in [n]$,
  and suppose
  $$\mathcal G - z\subseteq \mathcal P(V_1)\cup\cdots\cup\mathcal
  P(V_k)$$ for a partition $\{V_1,\ldots,V_k\}$ $(V_i\ne\emptyset,
  i=1,\ldots , k)$ of $[n]\setminus\{z\}.$ Then there exists $1\leq
  i\leq k$, let it be $i=1$, such that $z$ sees $V_1$, moreover, if it
  does not strongly see $V_1$, then $|\mathcal G(z)|\ge 2^{|V_1|}+ m-1
  ,$ where $m:=\min_{i=2,\ldots , k} |V_i|.$
\end{theorem}

Note that since $\mathcal G - z$ generates $[n]\setminus z$, in fact
the equality holds in the condition.  Introduce the notation $\mathcal
U:=\{U\subseteq V_1, \{z\}\cup U\notin \mathcal G\}.$ Then $z$ does
not strongly see $V_1$ if and only if $\mathcal U\ne\emptyset$;
$\{z\}\in\mathcal G$ implies $\emptyset\notin\mathcal U$, and
therefore $\mathcal U$ has a nonempty member which is inclusionwise
minimal.

\medskip
\begin{proof}
  Suppose for a contradiction that the first part of the theorem is
  false, that is, for all $i\in\{1,\ldots, k\}$, there exists
  $\alpha_i \in \mathcal P(V_i)\backslash (\mathcal G(z)\sqcap{V_i})$.
  Since $\{z\}\in\mathcal G$ we have $\emptyset\in\mathcal
  G(z)\sqcap{V_i}$, so $\alpha_i\ne\emptyset$ for all $i=1,\ldots, k$.

  Let now $Z:=\{z\}\cup\alpha_1\cup\cdots\cup\alpha_k$. The set $Z$ is
  generated by at most $k$ members of $\cal G$, exactly one of which,
  -- denote it by $g$ -- contains $z$. Clearly, $g\cap V_i\subseteq
  \alpha_i$, and $g\ne\alpha_i$ because of the definition of
  $\alpha_i$ $(i=1,\ldots, n)$. So $Z\setminus g$ still contains an
  element from each $V_i$ $(i=1,\ldots, k)$, and therefore cannot be
  generated by at most $k-1$ members of $\mathcal G - z\subseteq
  \mathcal P(V_1)\cup\cdots\cup\mathcal P(V_k)$. This contradiction
  proves the first part of the theorem. That is, we can now assume
  $G(z)\sqcap{V_1}=\mathcal P(V_1)$, define $\mathcal U$ like before
  the proof, and note: {\em if $g\in\mathcal G(z)$, $g\cap
    V_1=U\in\mathcal U$ then $g$ meets $[n]\setminus (V_1\cup z)$.}

  \medskip To prove the stronger inequality of the theorem, let $U\in
  \mathcal U$ be minimal in $\mathcal U$; as noted $U\ne\emptyset$.
  Define \\$\mathcal G_{=U}:=\{g\in\mathcal G(z): g\cap V_1 =
  U\} = \mathcal G(z\cup U)-(V_1\setminus U)$, and\\
  $\mathcal G_{\subsetneq U}:=\{g\in\mathcal G(z): g\cap V_1
  \subsetneq U\} = \mathcal (G(z)-(V_1\setminus U))\setminus \mathcal
  G_{=U}$. Clearly, $\mathcal G_{=U}\cap \mathcal G_{\subsetneq
    U}=\emptyset.$ Let $\tau:=\tau(\mathcal G_{=U}/(U\cup z)),$ that
  is, $\tau$ is the minimum size of a set disjoint of $U\cup z$ that
  meets each member of $\mathcal G_{=U}$. This minimum is finite,
  since as noted, each member of $\mathcal G_{=U}$ has an element
  outside $U$.  Note also that $|\mathcal H |\ge \tau(\mathcal H)$
  holds whenever the latter is finite. Therefore we can suppose
  $\tau<m$ without loss of generality, since otherwise $|\mathcal
  G_{=U}|\ge \tau\ge m,$ and
$$|\mathcal G(z)|=|\mathcal G(z)\setminus \mathcal G_{=U} |+ |\mathcal
G_{=U}|\ge (2^{|V_1|}-1)+m,\leqno{{\rm (ineq 1)}}$$ and nothing else
remains to be proved.

\bigskip \noindent {\bf Claim}: $|\mathcal G_{\subsetneq U}|\ge
2^{|U|}+ 2^{m-\tau} - 2$.

\medskip
Since $U\in\mathcal U$ is minimal, $z\sqcup (\mathcal P(U)\setminus
U)\subseteq \mathcal G_{\subsetneq U},$ so we know already
$2^{|U|}-1$ elements of $\mathcal G_{\subsetneq U}$. It suffices to
show now that $\mathcal G_{\subsetneq U}$ has at least $2^{m-\tau} -
1$ elements that meet $[n]\setminus V_1$.

\smallskip Let $C$ be a transversal of $\mathcal G_{= U}/(U\cup z)$,
$|C|=\tau$. Then $C\subseteq V_2\cup\ldots \cup V_k$. Now the
condition of the theorem is satisfied for $\mathcal G - ((V_1\setminus
U)\cup C)$, with the same $z$, and with the partition
$\{U,V_2\setminus C,\ldots, V_k\setminus C\}$: we already know
$U\ne\emptyset$, and because of $|C|=\tau<m$, $V_i\setminus
C\ne\emptyset,\,(i=2,\ldots,k)$.

Since $U\in\mathcal U$ and $C$ is a transversal of $\mathcal
G_{=U}/(U\cup z)$, $\mathcal G(z) - (V_1\setminus U)\cup C=\mathcal
G_{\subsetneq U}-C$. Since $z$ does not see $U$, by the already proven
first assertion of our theorem it does see $V_i\setminus C$ for some
$i=2,\ldots, k$.  Let $i=2$: $V_2\setminus C$ has at least $m-\tau$
elements, and therefore $\mathcal P(V_2\setminus C)$ has at least
$2^{m-\tau}-1$ non-empty members.

\medskip Using that $z$ sees $V_1$, and then applying the Claim and
the inequality $2^{m-\tau}\ge m-\tau+1$ we get:
$$|\mathcal G(z)|=|\mathcal G(z)\setminus (\mathcal G_{= U}
\cup \mathcal G_{\subsetneq U})| + |\mathcal G_{= U}| + |\mathcal
G_{\subsetneq U}|\ge 2^{|V_1|} - 2^{|U|} + \tau + 2^{|U|}+ 2^{m-\tau}
- 2\ge $$
$$\ge 2^{|V_1|} + \tau+ (m-\tau+1) - 2= 2^{|V_1|} + m -
1.$$
\end{proof}

The equality case of the bounds is worth analyzing also in hope of
gains in the estimates: the gains allow to deduce stronger bounds on
the degree from weaker bound, and therewith the optimality of
$\Constr(n,k)$ for some $n$ and $k$. In the following analysis and
corollary we will suppose $\mathcal G\subseteq \mathcal P ([n])$ is an
$(n,k)$-generator, $z\in [n]$, and $\mathcal G - z\subseteq \mathcal
P(V_1)\cup\cdots\cup\mathcal P(V_k)$ for a partition
$\{V_1,\ldots,V_k\}$ $(V_i\ne\emptyset, i=1,\ldots , k)$ of
$[n]\setminus\{z\};$ we denote $\mu,\, m$ the smallest and the second
smallest size among the sizes $\{|V_i| : i=1,\ldots , k \}$ of the
partition classes.

Under the condition of the theorem a first estimate is $|\mathcal
G(z)|\ge 2^\mu$, since $z$ sees one of the classes. The theorem claims
that there is equality in this bound if and only if $z$ strongly sees
one of the smallest classes.

It is interesting that the bound jumps from $2^{|V_1|}$ to
$2^{|V_1|}+m-1$ if $z$ sees $V_1$ but does not strongly see it.  What
are the conditions of the equality then ?

\begin{proposition}\label{prop:long} Suppose  $\mathcal G\subseteq \mathcal P ([n])$
  is an $(n,k)$-generator, $z\in [n]$, and $\mathcal G - z\subseteq
  \mathcal P(V_1)\cup\cdots\cup\mathcal P(V_k)$ for a partition
  $\{V_1,\ldots,V_k\}$ $(V_i\ne\emptyset, i=1,\ldots , k)$ of
  $[n]\setminus\{z\}.$ If $z$ sees $V_1$ but does not strongly see it,
  that is, $\mathcal U\ne\emptyset$, then the equality holds in the
  bound
$$|\mathcal G(z)|\ge 2^{|V_1|} + m - 1,\leqno{(\rm ineq 2)}$$
if and only if there exists $1\leq i\leq k$, let it be $i=2$ such that
$|V_2|=m$, $V_2=:\{v_1,\ldots,v_m\}$ and choosing the indices
appropriately, one of (i)-(iii) is true:
\begin{itemize}{\em \item[(i)] There exists $U\subseteq V_1$ such that
    with $\mathcal U_1:= (\mathcal P(U)\setminus\{U\})$ and $\mathcal
    U_2=\{U\cup\{v_i\}: i=1,\ldots,m\}$, or $\mathcal
    U_2=\{U\cup\{v_i\}: i=1,\ldots,m-1\}\cup \{\{v_m\}\},$ $$\mathcal
    G(z)/z=\mathcal U_1\cup \mathcal U_2.$$ \item[(ii)] $m=2$,
    $\mathcal U\subseteq \mathcal P(V_1)$ is arbitrary,
    $g_{=U}:=U\cup\{v_1\}$ $(U\in\mathcal U), and$
$$\mathcal G(z)/z=(\mathcal P(U)\setminus \mathcal U)\cup \{g_{=U}: U\in\mathcal U\}\cup \{v_2\}$$
\item[(iii)] $m=1$, $\mathcal U$ is arbitrary, and $\mathcal
  G(z)/z=(\mathcal P(U)\setminus \mathcal U) \cup \{g_{=U}:
  U\in\mathcal U\},$ where $g_{=U}$ is the union of $U$ and an
  arbitrary non-empty set of elements that form singleton classes.}
\end{itemize}
\end{proposition}

\begin{proof}
  Suppose the condition of Theorem~\ref{thm:Strong} is satisfied, and
  (ineq2) is satisfied with equality. Then $m\le \tau$, since $m>\tau$
  would imply that (ineq1) would also be satisfied with strict
  inequality, and then so would be the identical (ineq2). To have
  equality in the claim, $\mathcal G(z)$ cannot contain a set that
  meets a partition-class of size bigger than $m$ different from
  $V_1$.

  Consider now $\mathcal U$ as in the proof, and let $U\in\mathcal U$.
  Let us now exploit the equalities in the inequalities of the proof
  of (ineq2) in the proof of Theorem~\ref{thm:Strong} from the end
  backwards: in order to have equality in (ineq2), we need
  $2^{m-\tau}= m-\tau+1,$ and since $m-\tau\ge 0$, this holds if and
  only if $m-\tau =1$, or $m-\tau=0$. We will have to consider both
  the case $\tau=m$ and $\tau=m - 1$.

  If $m>2$ then $|\mathcal G_{=U'}|>1$ for all $U'\in\mathcal U$,
  while in (ineq1) we used the bound of $1$ for all but one
  $U\in\mathcal U$. So the strict inequality holds if $|\mathcal U |
  >1$. If $|\mathcal U | =1$ the equality can hold, and the two cases
  corresponding to $\tau=m$ and $\tau=m-1$ are listed in (i).

  If $m=2$ and $\tau=m$, then again, $|\mathcal G_{=U'}|>1$ for all
  $U'\in\mathcal U$, and the strict equality can hold only if
  $|\mathcal U | =1$, included already in the previous case. However,
  if $m=2$ and $\tau=m-1$, then $|\mathcal G_{=U'}|=1$ is possible for
  all $U'\in\mathcal U$, and precisely if the unique element of
  $\mathcal G_{=U'}$ is the $\mathcal G_{=U'}$ of (ii). So all the new
  cases where equality can occur for $m=2$ are listed in (ii).

  If $m=1$, then as noticed, all sets in $\mathcal G(z)$ must be
  included in the union of $V_1$ and the partition classes of size
  $m$, that is, must be of the form given in (iii). It is easy to
  check that this is then sufficient: all sets of this form are
  $(n,k)$-generators. \iffalse If there exists a transversal
  $C\subseteq V_2\cup\ldots\cup V_k$ of $\mathcal G_{=U}/(U\cup z)$
  such that for all $i=2,\ldots,k$ we have $|V_i\setminus C|>1$, then
  $\mathcal G_{\subsetneq U}$ has at least $4$ elements meeting
  $V\setminus (U\cup z)$, so at leat $3$ non-empty elements, and
  therefore !!!!!!!!  $|\mathcal G(z)|\ge |\mathcal G_{=U}/(U\cup z)|+
  3\ge $, so the equality cannot be satisfied in (ineq2). \fi
\end{proof}

\bigskip We get the
following corollary from  the theorem and the above analysis of the
equality.  Recall the notations $p$ and $m$.

\begin{corollary}\label{cor:sharp} $\mathcal G\subseteq \mathcal P
  ([n])$
   is an $(n,k)$-generator, $z\in [n]$,
  and
  $\mathcal G - z\subseteq \mathcal P(V_1)\cup\cdots\cup\mathcal
  P(V_k)$ for some partition $\{V_1,\ldots,V_k\}$
  $(V_i\ne\emptyset, i=1,\ldots , k)$ of $[n]\setminus\{z\}.$ Then
  $$| \mathcal G (z)
  | \ge 2^{p-1} + m\leqno{(ineq3)}$$ unless $z$ strongly sees one of
  the classes, or one of (i), (ii), (iii) holds.
\end{corollary}

\medskip The following lemma states in addition to the optimality of
the construction the unicity of optima -- that is,
Conjecture~\ref{conj:Strong} -- by induction on $n$ if and only if
every optimal generator contains a vertex of degree at least $2^{p-1}$
(which is still somewhat weaker than Conjecture~\ref{conj:Degree}, see
Conjecture~\ref{conj:StrongDegree}):

\begin{lemma}
  \label{lemma:StrongInduct}
  Let $\mathcal G$ be an optimal $(n,k)$-generator, $z\in [n]$,
  $|\mathcal G(z)|\ge 2^{p-1}$ and $p\ge 3$; assume that
  $\Constr(n-1,k)$ is the unique optimal $(n-1,k)$-generator. Then
  $\mathcal G=\Constr(n,k)$.
 \end{lemma}

 \begin{proof} By Lemma~\ref{lemma:WeakInduct}, $|\mathcal G(z)|=
   2^{p-1}$, and $|\mathcal G|=\constr(n,k)$, whence $\mathcal
   G-z=\constr(n,k)-2^{p-1}=\constr(n-1,k)$, and then by the
   condition, $\mathcal G - z = \Constr(n-1,k)$.

   So $\mathcal G - z= \mathcal (P(V_1)\cup\cdots\cup\mathcal P(V_k))\backslash\{\emptyset\}$,
   where $\{V_1,\ldots,V_k\}$ is a partition of $[n]$ into parts of
   size $p(n,k)$ and $p(n,k)-1$.  By Theorem~\ref{thm:Strong} one can
   choose $V_1$ so that either $\mathcal G(z)/z=\mathcal P(V_1)$, or
   $|G(z)|\ge 2^{|V_1|}+ m-1$ with $m=\min_{i=2,\ldots , k} |V_i|=p(n,k)-1\geq 2$.

   In the first case, by optimality, $V_1$ is a class of size
   $p(n,k)-1$ so that $\mathcal G=\Constr(n,k)$ follows. If
   indirectly, the second case holds, then 
$$2^{p-1}=|\mathcal G(z)|\geq 2^{p-1}+ m-1\geq 2^{p-1}+1,
$$
and this contradiction finishes the proof.
\end{proof}

\medskip
Modified as follows, Conjecture~\ref{conj:Degree} becomes equivalent
to Conjecture~\ref{conj:Weak} by Lemma~\ref{lemma:WeakInduct}.

\begin{conjecture}%[Weak Degree Conjecture]
  \label{conj:WeakDegree}
  For all $n,k\in\Nset$ there exists an optimal $(n,k)$-generator
  $\mathcal G$ and $z\in [n]$ such that:
 $$|\mathcal G(z) | \ge 2^{p(n,k)-1}.\leqno{(2)}$$
\end{conjecture}

Modified as follows, Conjecture~\ref{conj:Degree} becomes equivalent
to Conjecture~\ref{conj:Strong} by Lemma~\ref{lemma:StrongInduct}.

\begin{conjecture}%[Strong Degree Conjecture]
  \label{conj:StrongDegree}
  For all $n,k\in\Nset$, for every optimal $(n,k)$-generator $\mathcal
  G$ there exists $z\in [n]$ such that (2) holds.
\end{conjecture}

We have thus the following implication between the conjectures :

\noindent Conjecture~\ref{conj:Degree} $\implies$
Conjecture~\ref{conj:Strong} $\implies$ Conjecture~\ref{conj:Weak},

\noindent
Conjecture~\ref{conj:details} $\implies$ Conjecture~\ref{conj:Weak}

\noindent Conjecture~\ref{conj:Weak} $\iff$
Conjecture~\ref{conj:WeakDegree},

\noindent Conjecture~\ref{conj:Strong} $\iff$
Conjecture~\ref{conj:StrongDegree}.

\medskip Let us also state the conjecture asserting that the
disjointness requirement does not change the optimum value.

\begin{conjecture}
\label{conj:disjoint} For all $n,k\in\Nset$: $\op(n,k)=\opt(n,k)$.
\end{conjecture}

\

So far all the simple Propositions, Lemmas and Conjectures hold
without change if disjointness is not required and $\op$ is written
instead of $\opt$. This is not true though for
Theorem~\ref{thm:Strong} and its corollaries, including
Lemma~\ref{lemma:StrongInduct} and Proposition~\ref{prop:long}, the
reason being that we used in an essential way that at most one of the
$k$ disjoint sets contains a given $z\in[n]$.

\section{Case $p\le 3$}\label{sec:main}
Recall the notation $p=p(n,k)=\lceil \frac{n}{k} \rceil$ and $n=p\,
k-r$ with $0\le r<k$. In this section we prove all the conjectures for
$p\leq 3$. This is done in Theorem~\ref{th:p2} for $p\le 2$, and in
Theorem~\ref{th:degreep=3} for $p=3$. (In the Appendix we add to this the two
first cases with $p=4$: $(n,k)=(7,2)$ and $(n,k)=(8,2)$.)

\begin{theorem}
  \label{th:p2}
  If $p\le 2$, that is $1\leq n \leq 2k$, then
  $\op(n,k)=\opt(n,k)=\constr(n,k)$, furthermore, for any (not
  necessarily optimal) $(n,k)$-generator $\mathcal G$, (1) holds,
  and there exists $z\in [n]$ such that (2) holds. A generator $\mathcal G$ is
  optimal  if and only if it consists of all the singletons
  in $[n]$ and $n-k$ pairwise disjoint sets of size at least $2$.
\end{theorem}

In particular, the construction is the unique optimal generator if
$n\le k$ or $n=2k$, but it is not unique if $k< n <2k$.  However, if
$k< n <2k$, Conjecture~\ref{conj:details} follows still easily, and it
is also not an exception of Theorem~\ref{thm:Strong} or the
reformulation of its essential part in Lemma~\ref{lemma:StrongInduct},
useful for proving unicity; this case is an exception to unicity only
because for $m=1$ -- and only in this case -- Theorem~\ref{thm:Strong}
does not exclude other optimal solutions of the same size, and they
indeed, exist, and are already mentioned in the (iii) case of
equality. The reason for this is nothing more than the validity of
$2^{p-1} +m-1=2^{p-1}$ in this case.

This is also the only case when ``Tur\'an's bound'' $T(n,k+1,2)$ is
exact.

\medskip
\begin{proof} Let $\mathcal G$ be an arbitrary $(n,k)$-generator. It
  contains all the singletons, and if $p=1$, that is, $n\le k$ there
  is no need of more members.

  If $p=2$, that is, $k+1\leq n \leq 2k$, then by
  Proposition~\ref{prop:extreme}, $|\mathcal G^2|\ge
  \constr^2(n,k)=k-r=n-k,$ and the equality holds if and only if the
  sets of size at least $2$ are disjoint.

  Conversely, suppose the hypergraph $\mathcal G$ has $n-k$ disjoint
  members of size at least $2$ $(k+1\leq n \leq 2k)$, and let us check
  that it is an $(n,k)$-generator. Let $S\subseteq [n], s:=|S|>k$.
  Then $S$ misses at most $n-s < k$ members of $\mathcal G^2$, so it
  contains at least $n-k - (n-s)=s-k$ members of $\mathcal G^2$, all
  pairwise disjoint. So $S$ can be generated by $s-k$ members of
  $\mathcal G^2$ plus at most $s-2(s-k)=2k-s$ singletons.

\end{proof}

\begin{theorem}
  \label{th:p3}
  If $p=3$, that is $2k< n \leq 3k$, then for any (not
  necessarily optimal) $(n,k)$-generator $\mathcal G$, (1) holds.
\end{theorem}

\begin{proof}
  We have (1) for $i=3$ by Proposition~\ref{prop:extreme}: $|\mathcal
  G^3|\ge \constr^3 (n,k)= n-2k.$

  Now we prove (1) for $i=2$, by induction on $n-2k$. By
  Theorem~\ref{th:p2} it is true for $n=2k$.  For the sake of easier
  understanding, we first do the proof separately for $n=2k+1$, using
  it for $n=2k$: For all $z\in [2k+1]$ we have $|\mathcal G^2 -
  z|\ge\constr^2(2k,k)+1=k+1$, otherwise we are done by
  Lemma~\ref{lemma:StrongInduct}. Now
$$\sum_{z\in [n]} |\mathcal G^2 -
z|\ge (2k+1)(k+1),$$ and in this sum every member of $G^2$ is counted
at most $2k-1$ times, so $|\mathcal G^2|\ge \frac{2k+1}{2k-1}(k+1) =
\frac{k+1/2}{k-1/2}(k+1)> k+2$. (For an easier look at it we used here
that multiplying a number $x$ by $\frac{k+1/2}{k-1/2}$ it increases by
more than $1$ if and only if $x> k-1/2$.)  Since
$\constr^2(2k+1,k)=\constr^2(2k,k)+3,$ $|\mathcal G^2|\ge
k+3=\constr^2(2k,k)+3=\constr^2(2k+1,k)$, as claimed.

Similarly, for an arbitrary $(n,k)$-generator, $2k+1\le n\le 3k$, we
have
$$|\mathcal G^2|\ge
\frac{n}{n-2}(\opt(n-1,k)-(n-1)+1) > \opt(n-1,k)-(n-1)+2,$$ since
$\opt(n-1,k)>\frac{n-2}{2}$, and the statement follows then using
$\constr^2(n,k)=\constr^2(n-1,k)+3$.
\end{proof}

We do not see how to deduce Conjecture~\ref{conj:Degree} from the
above theorem.  On the other hand, we can prove this conjecture
separately (for $p=3)$, implying the previous theorem as well, in a
simpler way, and without using any of the previous results or the
disjointness of generators. (For $i=3$ (2) is easy, and the following
theorem implies it for $i=2$ and $i=1$. For $2k\leq n \leq 3k$ we will
thus have two proofs of the optimality. (We still included
the previous theorem because it forecasts our future difficulties:
whenever the average degree of $\Constr(n,k)$ is much smaller than the
maximum degree, ``averaging arguments'' do not easily work.)

\begin{theorem}
  \label{th:degreep=3}
  If $p=3$, that is $2k< n \leq 3k$, then
  $\op(n,k)=\opt(n,k)=\constr(n,k)$, furthermore, for any (not
  necessarily optimal) $(n,k)$-generator $\mathcal G$, (1) holds, and
  there exists $z\in [n]$ such that (2) holds. The construction is the
  unique $(n,k)$-generator.
\end{theorem}

\begin{proof}
  We prove, without requiring disjointness, that for any
  $(n,k)$-generator $\mathcal G$, there exists $z\in [n]$ such that
  (2) holds.

  We can suppose without loss of generality $n=2k+1$.  Indeed, if
  $n>2k+1$, then we can apply the proven assertion to the
  $(2k+1,k)$-generator $\mathcal G(U)$, where $U\subseteq[n]$,
  $|U|=2k+1$.

  Let $\mathcal G$ be an $(n,k)$-generator, and suppose for a
  contradiction $|\mathcal G^2(z)|\le 2$ for all $z\in[n]$.

  We define an undirected graph $G=(V,E)$ on $V:=[n]=[2k+1]$, in the
  following way: for each $g\in\mathcal G$, $|g|\geq 2$, we choose two
  vertices $u,v\in g$, let $e=uv\in E$, and use the notation $g_e$ for
  $g$.  For $g_1\ne g_2\in\mathcal G$ we can take the same $u, v$ (if
  $u,v\in g_1\cap g_2),$ but then we take two parallel $uv$ edges
  $e_1$ and $e_2$. We will say that the edge $e=uv$ {\em represents}
  $g_e\in\mathcal G$. We thus suppose that different sets in $\mathcal
  G$ are represented by different edges.  Furthermore, we suppose that
  we make the possible choices of $u$ and $v$ so as to minimize the
  number of components of $G$.

  Now it follows from the indirect assumption that all the degrees of
  the graph $G$ are at most $2$, so it is a disjoint union of cycles,
  paths and isolated vertices. The following Claim is the key of the
  proof:

\medskip
\noindent {\bf Claim}: Let $C$ be a cycle of $G$, and $e$ an edge of
$C$. Then $e\in\mathcal G$, and is not contained in any bigger set of
$\mathcal G$.

\medskip
Indeed, by the definition of $G$, $e$ is contained in a set of
$\mathcal G$, so it is sufficient to prove  that no set in $\mathcal
G$ can properly contain $e$.

\begin{itemize}
\item[--] If an extra element $z$ of $g_e$ (different from the
  endpoints of $e$) of such a set were in $C$, then $z$ would be
  contained in three different sets of $\mathcal G$: $g_a$ and $g_b$,
  where $a,b$ are the two edges incident to $z$ in $C$, and $g_e
  \supseteq e\cup\{z\}$.  Clearly, $e$, $a$, $b$ are different, and
  therefore $g_e$, $g_a$, $g_b$ as well, contradicting the indirect
  assumption.
\item[--] If an extra element $z$ of $g_e$ (different from the
  endpoints of $e$) of such a set were in another component $K$ of
  $G$, then replacing one of the endpoints of $e$ by a point in
  $g_e\cap K$, we get another representation of $\mathcal G$ with one
  less component (all vertices of $C$ and $K$ are now in the same
  component), contradicting the definition of $G$.
\end{itemize}
The claim is proved.

Let $U$ be the set of vertices of $G$ that are in a cycle.  The
subgraph $G(V\setminus U)$ contains only paths and isolated vertices,
so we can find a stable set (not containing both endpoints of an edge)
$S$ of $G(V\setminus U)$ such that $|S|\geq |V\setminus U|/2$. (We
take a (the) bigger stable set in each component.)

We show now that $S\cup U$ cannot be $k$-generated, contradicting the
choice of $\mathcal G$.  Recall that any $g\in\mathcal G$, $g\subseteq
S$ has also an edge in $G$. But the only edges in $S\cup U$ are in the
cycles, and for these the claim holds. Therefore what we have to show
is exactly that $S\cup U$ is not the union of at most $k$ edges of $G$
or singletons.

Indeed, denote $\gamma (X)$ the minimum number of edges and singletons
necessary for generating a set $X\subseteq n$. Let the components of
$G$ be $C_1,\ldots, C_t$ $(t\in\Nset)$. Note that for all $i=1,\ldots,
t:$ $\gamma(U\cap C_i)\ge |C_i|/2$. Then
  $$\gamma(U)=\sum_{i=1}^t \gamma(U\cap C_i)\ge \sum_{i=1}^t |C_i|/2
  =\frac{2k+1}2 >k.$$

So $U$ cannot be $k$-generated, a contradiction. 

By lemma~\ref{lemma:WeakInduct} (that does not require
disjointness), $\op(n,k)=\opt(n,k)=\constr(n,k)$ follows.

When disjointness is required, by lemma~\ref{lemma:StrongInduct}, the
contruction is the unique optimal $(n,k)$-generator.
\end{proof}

\section{Optimization and approximation}
\label{sec:opt}

The general problem this work is concerned with is natural to be asked
in terms of combinatorial optimization, including also computational
complexity and approximation ratios. In this section we would like to
present our related observations: some negative results concerning the
computational complexity, and simple but surprisingly good estimates
for the quantity $\opt (n,k)$.

Two natural optimization problems arise:

\begin{itemize}
\item[--] We do not want to generate all cars, that is, all subsets of
  options, just a pre-given family.

\item[--] The generator is restricted to choose elements from a given
  hypergraph.
\end{itemize}

More precisely:

\bigskip
\noindent {\bf PROBLEM:} CHOOSY CUSTOMER'S DIVERSITY\\
{\bf Input}: $\mathcal C\subseteq \mathcal P([n])$, numbers $k, s$.\\
{\bf Question}: Does there exist $\mathcal G\subseteq \mathcal P([n])$
that $k$-generates all sets in $\mathcal C$, and $|\mathcal G|\le s.$

\bigskip
\noindent {\bf PROBLEM:} CONSTRAINED PRODUCER'S DIVERSITY\\
{\bf Input}: $\mathcal H\subseteq \mathcal P([n])$, number $k$ and a target-set $T\subseteq [n].$\\
{\bf Question}: Does there exist $\mathcal G\subseteq \mathcal H$ that
$k$-generates $T$ ?

\medskip Note that in this second problem we only speak about the {\em
  existence} of a generator. These are just two simple and natural
variants that we choose for the sake of examples. The reader may enjoy
stating his favorite variants and checking NP-completeness for them.

\begin{theorem} Both CHOOSY CUSTOMER'S and CONSTRAINED PRODUCER's
  DIVERSITY problems are NP-complete.
\end{theorem}

\begin{proof}
  We first reduce VERTEX COVER to CHOOSY CUSTOMER'S DIVERSITY, and
  even to instances where $k=2$. (VERTEX COVER and 3DM below are
  proved to be NP-complete in Garey and Johnson's seminal book
  \cite{GJ}.)

  Let $G=(V,E)$ be a graph, and consider the problem with input
  $\Omega=V\cup \{u\}$, where $u$ is an extra vertex not in $V$, and
  ${\cal C}:=\{\{v\}:v\in \Omega\}\cup \{ \{a,b,u\} : a,b\in V, ab\in
  E \}$.

  Clearly if $T$ is a vertex cover, that is $T\cap e\ne\emptyset$ for
  all $e\in E$, then $\mathcal G:=\{\{v\}:v\in \Omega\}\cup\{\{t, u\}
  : t\in T\}$ does $2$-generate all $C\in\mathcal C$. Conversely,
  $\{\{v\}:v\in \Omega\}$ must be contained in all generators, and all
  the other sets can be supposed to contain $u$ and to be of size $2$.
  (Otherwise we can add $u$ and keep only one of the elements
  different from $u$.) Let $T:=\{v\in V: (v,u)\in\mathcal G\}$. Then
  $T$ is a vertex cover, finishing the proof of the first assertion.

  Let us now reduce 3DM to CONSTRAINED PRODUCER'S DIVERSITY. Let
  $(U,V,W, E)$ be an instance of 3DM, that is, $E\subseteq U\times
  V\times W$ (the Cartesian product of $U, V, W),$ where
  $|U|=|V|=|W|=3k$. Define $T:=U\cup V\cup W$. Now clearly, $\mathcal
  G \subseteq E$ $k$-generates $T$ if and only if it is a
  $3$-dimensional matching (that is, if and only if it partitions
  $T$).
\end{proof}

In both proofs it is irrelevant whether we ask disjointness or not
from the generators. (In these cases there exists always a disjoint
optimal solution.)

We now show that the construction provides a quite good approximation
of the optimum. Enumeration provides the bound $\constr(n,k)\le
\opt(n+2k,k)$. Let us sketch a proof of this. Given an $(n,k)$
generator $\mathcal G$, all the $2^n - 1$ nonempty subsets of $[n]$
can be encoded by an at most $k$ element subset of $\mathcal G$:
$$\sum_{i=1}^{k}\binom{|\mathcal G|}{i}\geq 2^n-1.$$
It follows that $k|\mathcal G|^k /k!\,\ge 2^n$, that is, $|\mathcal
G|^k \ge (k-1)!\, 2^n$, and applying Stirling's formula and taking
the $k$-th root: $|\mathcal G|  \ge \frac{k-1}{e} 2^{n/k}$. So
$\opt(n,k) \ge \frac{k-1}{e} 2^{n/k}$, while $\constr(n,k)\le k
2^{n/k} +$const, which shows that $\constr(n,k) /\opt (n,k)$ does
not exceed $\varepsilon(n,k)e$ where
lim$_{n,k\rightarrow\infty}\varepsilon(n,k)=1$. The exact treshold
valid for all $n$ and $k$ is certainly smaller than $4$:
$\constr(n,k)\le 4\, \opt (n,k).$ Since $\constr(n+2k,k)\ge 4
\constr(n,k)$, we got that $\constr(n,k)\le \opt(n+2k,k)$.

For small $k$ we do not have to apply Stirling formula and we get
essentially better bounds: for $k=2$, we get $|\mathcal G | +
\binom{|\mathcal G|}{2} \geq 2^n-1$ and we get the same bounds as in
the theorems below. Still with the same method, for $k=3$ we get that
the construction is at most $\frac 4{\sqrt[3]{12}}=1,747\cdots$ times
the optimum. Let us deduce the results for $k=2$ with another method
as well, which will also lead to a simple general proposition for
arbitrary $k$:

\begin{theorem}\label{thm:bound2}
For all $n\in\Nset$:
$$\opt(n,2)\le \constr (n,2) \le 3/2 \opt (n,2),$$ and the constant 3/2
can actually be improved to $\sqrt 2$ if $n$ is even.
\end{theorem}

Expressing $\opt(n,2)$: $C\constr(n,2)\le\opt(n,2)\le\constr(n,2),$
with $C=2/3$ if $n$ is odd, and $C=\sqrt{2}/2$ if $n$ is even.

\bigskip
\begin{proof} Let $\mathcal G$ be an  $(n,2)$-generator. Since every
subset of $[n]$ containing $z$ is the union of a set in $\mathcal
G(z)$ and a set in $(\mathcal G - z)\cup\{\emptyset\}$, we have :
$$|\mathcal G (z)| (|\mathcal G - z|+1)\ge 2^{n-1}.$$
The minimum of $x+y$, $(x,y\in\Rset)$ under the condition $xy=2^{n-1}$
is $x=y=2^\frac{n-1}{2}$. Therefore, if in addition $\mathcal G$ is an
optimal $(n,k)$-generator, then
$$\opt(n,2) = |\mathcal G|= |\mathcal G (z)| + |\mathcal G - z|
\ge \min\{x+y-1: xy=2^{n-1}\} = 2^{\frac {n-1} {2}} + 2^{\frac {n-1}
{2}}-1.$$ On the other hand, $\constr(n,2)= 2^{\frac {n-1} {2}}-1 +
2^{\frac {n+1} {2}}-1$ if $n$ is odd, and $\constr(n,2)= 2^{\frac
{n} {2}}-1 + 2^{\frac {n} {2}}-1$ if $n$ is even.
\end{proof}

If we compare $\constr(n,2)$ with the same estimates applied to
$\opt(n+1,k)$ or $\opt(n+2,k)$, we get the following:

\begin{theorem}\label{cor:bound2}
  For all $n\in\Nset$: $\opt(n,2)\le \constr (n,2) \le \opt
  (n+1,2),$ if $n$ is even, and $\opt(n,2)\le \constr (n,2) \le
  3/4\opt (n+2,2),$ if $n$ is odd.

\end{theorem}

\medskip Finally, we prove now with the same method a general
statement which has only self-interest so far: for a hypergraph
$\mathcal H$ let
$$\alpha (\mathcal H):=\max\{S\subseteq [n]: \hbox{$H\cap S$ is a singleton for all $H\in \mathcal H$}
\}.$$

Note that an $(n,k)$-generator $\mathcal G$ always satisfies $\alpha
(\mathcal G)\le k$.  On the other hand, for all $n,k$,
$\alpha(\Constr(n,k))=k$.  Conversely, a generator $\mathcal G$ with
$\alpha (\mathcal G)=k$ looks close to the optimum, and we can easily
prove that it is optimal, if $n=pk$:

\begin{proposition}\label{prop:pairs} Suppose $n=pk$,
  $\opt(n-k,k)=constr(n-k,k)$, and that there exists an optimal
  $(n,k)$-generator $\mathcal G$, $\alpha (\mathcal G)=k$. Then
  $opt(n,k)=\constr(n,k)$ and $\Constr(n,k)$ is the unique
  optimal $(n,k)$-generator, provided the same holds for $(n-k,k)$.
\end{proposition}

If $k=2$, the condition is $\alpha (\mathcal G)=2$ and this means
$x,y\in [n]$ such that $\mathcal G(x)$ and $\mathcal G(y)$ have no
common elements. The proposition confirms all the conjectures under
this condition (which is true for $\Constr(n,2))$.

Let $S$ be a set that meets all members of $\mathcal H$ only in one
element, $|S|=k$. We will actually show that at least $\constr(n,k)$
sets are needed {\em only to generate all sets in $S\sqcup
  P([n]\setminus S)$ and in $\mathcal P([n]\setminus S)$} !

\medskip
\begin{proof} Clearly, any set containing $S$ is generated by exactly
  $k$ sets, exactly one from each $\mathcal G(s)$ $(s\in S).$ Thus
  $$\prod_{s\in S} |\mathcal G(s)| \ge 2^{n-k}.$$ By the inequality
  between the geometric and arithmetic means, we have under this
  condition $$\sum_{s\in S} |\mathcal G(s) | \ge k 2^{\frac{n-k}k}=
  k2^{p-1}.\leqno{(ineq4)}$$

  The equality holds in (ineq4) if and only if $|\mathcal
  G(s)|=2^{\frac{n-k}k}$, and the members of $\cup_{s\in S} \mathcal
  G(s)$ generate $\prod_{s\in S} |\mathcal G(s)|$ sets; the latter
  condition holds if and only if any pair of sets from different
  $\mathcal G(s)$ are disjoint.

  Define for all $s\in S$, $P_s:=\cup \mathcal G(s)$. Because of
  $|\mathcal G(s)|=2^{\frac{n-k}k}$ we have $|P_s\setminus\{s\}|\ge
  \frac{n-k}k,$ that is,
$$\sum_{s\in S} |P_s|\ge k (\frac{n-k}k + 1)\ge n,$$
and if there is equality in (ineq4) and therefore the sets $P_s$ are
pairwise disjoint, then there is equality everywhere, that is,
$|P_s|=\frac{n-k}k + 1= n/k=p$ for all $s\in S$.

We have arrived now to our final estimation one ingredient of which is
(ineq4), and the other is the obvious inequality $|\mathcal G - S|\ge
\opt(n-k,k)$.  Then
$$\opt (n,k)=|\mathcal G| = |\mathcal G - S | + \sum_{s\in S} |\mathcal
G(s) |\ge \opt(n-k,k) + k2^{p-1}= $$

$$=  \constr(n-k,k) + k 2^{p-1} = \constr(n,k).$$ 
So $\opt(n,k)=\constr(n,k)$, and there is equality everywhere, so $\mathcal
G - S$ is optimal. If $\Constr(n-k,k)$ is the unique optimal
$(n-k,k)$-generator then $\mathcal G - S$ is isomorphic to
$\Constr(n-k,k)$.  Finally, applying Lemma~\ref{lemma:StrongInduct}
$k$ times one by one to the elements of $s$ in the role of $z$, we see
that $\mathcal G=\Constr(n,k)$.
\end{proof}

%We suspect that the unicity can be deduced here without supposing it
%by recursion for $(n-k,k)$, and without using
%Lemma~\ref{lemma:StrongInduct}.

\bigskip 
\noindent {\bf Conclusion}: We proved that the most natural
construction for an $(n,k)$-generator is optimal if $n\leq 3k$, and
for some other individual pairs $(n,k)$, regardless whether the
disjointness of the sets is required, moreover, it is always a
constant time approximation with a small constant. The natural
formulations as an optimization problem are NP-hard.

\section*{APPENDIX: $k=2$ and can we go further ?}\label{sec:two}

We deduce the conjecture for two more cases, also in order to provide
another example of applying the arguments and assertions of the paper,
and to realize the limits of some arguments.

\medskip The following lemma extends the validity of
Theorem~\ref{thm:Strong} to the case when $\mathcal G - z$ can
contain one more set besides subsets of the partition-classes. We
restrict ourselves to the case $k=2$ (the statement and its use seem
to be considerably more complicated (even if not hopeless) for $k>2$)
:

\begin{lemma}
  \label{lemma:SuperStrong}
  Suppose $\cal G$ is an (n,2)-generator, $(\mathcal G-z)\subseteq
  \mathcal P(V_1)\cup\mathcal P(V_2)\cup \{h\}$, where
  $\{\{z\},V_1,V_2\}$ $(z\in [n])$ is a partition of $[n]$, $2\le
  \mu:=|V_1|\le |V_2|$ $(i=1,2)$, $h\subseteq V$.  Then $| \mathcal
  G(z) |\geq 2^{\mu}$, in particular, $\mathcal G$ is not optimal.
\end{lemma}

Of course, we can suppose without loss of generality $h\cap
V_i\ne\emptyset$ $(i=1,2)$, otherwise $h$ can be omitted from
$\mathcal G$, and the assertion follows from
Theorem~\ref{thm:Strong}.

  \medskip
\begin{proof} If $z$ sees $V_1$ or $V_2$ we are done, so we suppose
it does not.

\smallskip\noindent {\bf Claim}: For both $i=1$ and $i=2$, there is at
most one subset of $V_i$ that is not in $\mathcal G(z)\sqcap V_i$.

\smallskip Suppose for a contradiction that the statement does not
hold say for $i=2$: let $B\ne C\subseteq V_2$, $B, C\notin \mathcal
G(z)\sqcap V_2$. Since $z$ does not see $V_1$, there exists
$A\subseteq V_1$, $A\notin\mathcal G(z)\sqcap V_1$. We show then
$|\mathcal G(z)|\ge 2^{|V_1|}$.

The sets $\{z\}\cup A\cup B$, $\{z\}\cup A\cup C$ must contain $h$
that must participate in $2$-generating these sets, whence
$$\{z\}\cup (A\cup
B)\setminus \{h\}, \{z\}\cup (A\cup C)\setminus \{h\} \in \mathcal
G.$$ We show now that $|\mathcal G(z)|\ge 2^{V_1}$, by labelling each
subset of $V_1$ with a different set in $\mathcal G(z)$.

If $U\subseteq V_1$, $U\in\mathcal G(z)\sqcap V_1$, we label $U$ with
an arbitrary $g\in\mathcal G(z)$, $g\cap V_1=U$. For instance we label
$\emptyset$ with $\{z\}$. If $A\notin\mathcal G(z)\sqcap V_1$, we saw
that there exist two sets, $\{z\}\cup (A\cup B)\setminus \{h\},
\{z\}\cup (A\cup C)\setminus \{h\}\in\mathcal G$.  At most one of them
is the label of $A\setminus \{h\}$, the other, say $(A\cup C)\setminus
\{h\}$ is a priori not a label, since it meets $V_1$ also in
$A\setminus \{h\}$, but it is not the label of this set.  Let the
label of $A$ be $(A\cup C)\setminus \{h\}$. Clearly, the label of a
different set $A'\subseteq V_1$, $A'\notin\mathcal G(z)\sqcap V_1$ is
different, since it is $A'\cup C\setminus \{h\}$, different from
$A\cup C\setminus \{h\}$.  (Both $A\cup C$ and $A'\cup C$ contain
$h$.)  The claim is proved.

\medskip The claim implies that $|\mathcal G(z)|\ge 2^{|V_1|}-1$, but
we are still fighting for the strict inequality here. Let $1\in h\cap
V_1$, $2\in h\cap V_2$. By Theorem~\ref{thm:Strong}, $z$ sees
$V_1\setminus\{1\}$ and $V_2\setminus\{2\}$ (since it does not see
$V_2$ and $V_1$). If it strongly sees both of them, then $z\sqcup
(V_1\setminus\{1\}), z\sqcup (V_2\setminus\{2\})\subseteq \mathcal G,$
and the only common element of these two is $z$, so the bound is
largely satisfied.  If not, then in Theorem~\ref{thm:Strong} the
equality is not satisfied, so there exists $A\subseteq V_1$ and
$f,g\in\mathcal G$ such that $A=f\cap V_1=g\cap V_1$ for $f\ne
g\in\mathcal G$, so the equality $|\mathcal G(z)|= 2^{|V_1|}-1$ does
not hold.
\end{proof}

\begin{theorem}
  \label{th:n7} For any (not
  necessarily optimal) $(7,2)$-generator, (1) holds, 
and  $\Constr(7,2)$ is the unique optimal generator.
\end{theorem}

\begin{proof} We first prove the second assertion.  Let $\mathcal G$
  be an optimal $(7,2)$-generator. Then $|\mathcal G|\leq
  \constr(7,2)$.  Add to $\mathcal G$ some new sets to get a
  hypergraph $\hat{\mathcal G}$ with $|\hat{\mathcal
    G}|=\constr(7,2)=22$. Obviously $\hat {\mathcal G}$ is still a
  generator. It suffices to prove now that $\hat{\mathcal
    G}=\Constr(7,2)$. Indeed, then $\Constr(7,2)=\hat{\mathcal
    G}=\mathcal G$ follows since $\hat{\mathcal G}$ does not contain
  any other generator properly.

Let $d:=1/n\sum_{x\in [n]} \hat{\mathcal G}(x)$ be the {\em average
degree} of $\mathcal G$. Clearly (as before, see
Proposition~\ref{prop:conjimpliesconj}):
$$dn=\sum_{x\in [n]}|\hat{\mathcal G}(x)|=\sum_{g\in\hat{\mathcal
    G}}|g|=\sum_{i=1}^n \hat{\mathcal G^i}.\leqno{(ineq5)}$$

\noindent {\bf Claim 1}:  $d>6$

\smallskip We already know $|\hat{\mathcal G}^1|\ge 22$ and therefore
$|\hat{\mathcal G}^2|\ge 15$ as well. At the other extreme
$|\hat{\mathcal G^4}|\ge 1$ is obvious, let $A\in \hat{\mathcal G^4}$.
We show $|\hat{\mathcal G}^3|\ge 5.$

\begin{itemize}
\item[--] If there exists $z\notin A$, $| \hat{\mathcal G}^3(z)|\ge 2$, then
  apply Proposition~\ref{prop:tau} after deleting $z$: $|\hat{\mathcal
    G}^3-z|\ge \tau(\hat{\mathcal G}^3-z ) \ge 2$. But this bound is
  self-improving: $|A|\ge 4$, so $A$ is not disjoint of the other set
  in $\hat{\mathcal G}^3-z$, and therefore $|\hat{\mathcal G}^3-z|\ge
  \tau(\hat{\mathcal G}^3-z )+ 1\ge 3$. But then $|\hat{\mathcal
    G}^3(z)|+ |\hat{\mathcal G}^3-z|\ge 2 + 3=5.$
\item[--] If there exists $z\in A$, $| \hat{\mathcal G}^3(z)|\ge 3$, then
  similarly, apply simply $|\hat{\mathcal G}^3-z|\ge
  \tau(\hat{\mathcal G}^3)-z \ge 2$ to get $|\hat{\mathcal G}^3(z)|+
  |\hat{\mathcal G}^3-z|\ge 3+ 2=5.$
\item[--] One of the preceding cases holds, because otherwise every
  $z\in[n]$ is covered by at most one member of $\hat{\mathcal
    G}^3\setminus \{A\}$, although there are at least $3$ sets of size
  at least $3$ in this hypergraph on $7$ elements.
\end{itemize}

We conclude now the proof of the claim by (ineq5):

  $$d\ge \frac{22 + 15 + 5 + 1}7 =
  \frac{43}{7} >6.$$

  According to the claim there exists $x\in [n]$, $|\hat{\mathcal
    G}(x)|\ge 7$, that is, $|\hat{\mathcal G}-x|\le 22 -
  7=15=\constr(6,2)+1$. If the strict inequality holds, we are done by
  Lemma~\ref{lemma:StrongInduct}, so we can suppose $|\hat{\mathcal
    G}(x)|= 7$.

  Now Proposition~\ref{prop:conjimpliesconj} can be applied for $n=6,
  k=2, p=3$: there exists $z\in [n]$, $\hat{\mathcal G}(z)-x\ge 2^{p-1} +1$.
  So $|\hat{\mathcal G}-\{x,z\}|\le\constr(5,2)$, and the equality holds
  here by Theorem~\ref{th:p2}.  Now Theorem~\ref{thm:Strong} can be
  applied to deduce that $z$ strongly sees the class of size $2$ of
  $\hat{\mathcal G} -\{x,z\}$, since $m=3$. So $\hat{\mathcal G} -x$ contains a
  hypergraph isomorphic to $\Constr (6,2)$, meaning that it is exactly
  $\Constr (6, 2)$ and one more element $h$. We conclude now the
  second part of the theorem with Lemma~\ref{lemma:SuperStrong}
  substituting $z$ for $x$.

  \medskip Let now $\mathcal G$ be an arbitrary $(7,2)$-generator. By
  the already proven part we have (1) for $i=1$ and $i=2$. It is also
  obvious for $i=4$; as above, denote $A\in\mathcal G^4$.  In exactly
  the same way as we proceeded above, we can get $|\mathcal G^3|\ge
  5$, after which it is still possible to do one more self-improving
  step, to prove $|\mathcal G^3|\ge 6=\constr^3(7,3)$, as claimed:

  Suppose for a contradiction $|\mathcal G^3|\le 5$. A set
  $T\in\mathcal G$, $|T|=3$ will be called a {\em triangle}.

\medskip
\noindent {\bf Claim 2}: If $|\mathcal G^3(z)|\ge 3$ then $\mathcal
G^3-z$ has exactly two disjoint triangles, and these partition
$[n]\setminus z$.  \medskip

\medskip Indeed, $|\mathcal G^3-z|\ge \tau (\mathcal G^3-z)\ge 2,$ and
if one of these two inequalities is strict, then we arrive at the
contradiction $5\ge |\mathcal G^3|= |\mathcal G^3(z)|+|\mathcal
G^3-z|\ge 3 + 3 =6.$

\bigskip The average degree of $\mathcal G^3$ is at least $\frac{4 + 3
  + 3 + 3 + 3}7= 16/7>2.$ So there exists $z\in\mathcal G$, $|\mathcal
G^3(z)|\ge 3$, and Claim 2 can be applied. Let $T_1$ and $T_2$ be the
two triangles of $G^3-z$ provided by Claim 2. Since $A$ is not a
triangle, it does not coincide with any of these, so $z\in A$. Let
$T_3\ne T_4\in \mathcal G(z)\setminus \{A\}$.

\medskip
\noindent {\bf Claim 3}: $T_3\cap T_4=\{z\}$.

\medskip Indeed, another common element of $T_3$ and $T_4$, denote it
by $x$, would also be contained in $T_1$ or $T_2$, say $T_1$. Then
$x\in T_1\cap T_3\cap T_4$, and also $x\in A$, since if not, $A$ with
$x\notin A\in\mathcal G^3$ is not a triangle, contradicting Claim 2.
But then $|\mathcal G^3(x)|\ge 4$, $|\mathcal G^3 - x|\ge \tau
(\mathcal G^3 - x)\ge 2$, contradicting the assumption $5\ge |\mathcal
G^3| (\ge 4 +2=6)$.

\bigskip It follows that $A$ meets one of $T_3$ and $T_4$ and not only
in $z$. Indeed, $$|A\setminus \{z\}|+ |T_3\setminus \{z\}| +
|T_4\setminus \{z\}|\ge 3 + 2 + 2=7>6=|[n]\setminus\{z\}|.$$ Let this
element be $x\in A\cap T_3 \setminus \{z\}$; since $x\in T_1\cup T_2$,
we can assume for instance $x\in T_1$.

Now again, Claim~2 can be applied to $x$, and since $|\mathcal G^3|\le
5$, both triangles of $\mathcal G - x$ are already among the listed
sets. These can be only $T_2$ and $T_4$, in particular $T_4$ is also a
triangle.

So $T_4=(T_1\setminus\{x\})\cup \{z\}$. Because of Claim~3, $T_3$
contains, besides $x\in T_1$ also an element of $T_2.$ Finally,
$A=\{x,z\}\cup (T_2\setminus T_3)$, since any other element in $A$
would again contradict Claim~2. It follows that $\mathcal G^4=\{A\}$.
In order to $2$-generate $\{1,2,3,4,5,6,7\}$ itself, we need a set in
$\mathcal G^4$ and its complement. But the complement of $A$ is
different of all of $T_1, T_2, T_3, T_4$, so $\mathcal G^3$ has a
sixth element, and this final contradiction finishes the proof of the
theorem.
\end{proof}

\begin{corollary}
  \label{th:n8}
   For any (not
  necessarily optimal) $(8,2)$-generator, then (1) holds,
  there exists $z\in [n]$ such that (2) holds, and $\Constr(8,2)$
is the unique optimal $(8,2)$-generator.
\end{corollary}

\begin{proof} $|\mathcal G^4|\ge 2$ is obvious as usually (since each
  $7$ element set still contains $g\in\mathcal G$, $|g|\ge 4$).
$$\sum_{z\in[n]}|\mathcal G^3 - z|\ge 8\, \hbox{$\constr^3$}(7,2)=48,$$
and every set of $\mathcal G^3$ has been counted at most $5$ times
in this sum, so $|\mathcal G^3|\ge \lceil 48/5 \rceil =10=\constr^3
(8,2)$.

It is now easy to prove $|\mathcal G^2|\ge \constr^2 (8,2)$ (and the
same $|\mathcal G^1|\ge \constr^1 (8,2)$), with the same argument as
in the proof of the previous theorem: it suffices to proof
$\hat{\mathcal G}$ with $|\hat{\mathcal G}|=\constr(8,2)=30$ is
nothing else but $\Constr (8,2)$, and for this it suffices to prove
that $\hat{\mathcal G}$ has an element of degree $2^3=8$. So the only
remaining assertion to prove is that for any $(8,2)$-generator with
$|\mathcal G|=\constr(8,2)=30$ there exists $z\in [n]$ such that (2)
holds. Then the last assertion also follows by
Lemma~\ref{lemma:StrongInduct}. Let $\mathcal G$ be such an
$(n,k)$-generator.

Let $d:=1/n\sum_{x\in [n]} \hat{\mathcal G}(x)$ be the {\em average
  degree} of $\mathcal G$. Clearly (as before, see
Proposition~\ref{prop:conjimpliesconj}):
$$d=1/n\sum_{x\in [n]}| \mathcal G(x)|=1/n\sum_{g\in\hat{\mathcal
G}}|g|=1/n\sum_{i=1}^n |\mathcal G^i|=1/8(30 + 22 + 10 +
1)>7,$$ finishing the proof of the corollary. \end{proof}

Note that for this last statement a much weaker bound is sufficient,
namely the first easy estimate of $|\mathcal G^3|$ without the
worksome one.

\bigskip In $\Constr(8,2)$ the average degree is equal to the maximum
degree and the same could be proved for the optimum generator, that is
why Corollary~\ref{th:n8} includes Conjecture~\ref{conj:Degree}. The
same can be proved for arbitrary even $n$ and $k=2$, but the odd $n$
case with ``small'' average degree remains open.

\bigskip
\noindent {\bf Aknowledgment}: \ We are indebted to Nicolas Trotignon
for useful discussions, among them to have noticed the variety of
optimal generators when $p=2$.  We also thank Zolt\'an F\"uredi for
connecting us to the current state of the subject.

\end{document}